\documentstyle[twocolumn,aps,pre,epsfig]{revtex}

\begin{document}
\draft
\preprint{Conformation of semiflexible polymer}
\title{Conformational transitions of a semiflexible polymer in nematic solvents}
\author{Akihiko Matsuyama\cite{byline}}
\address{Department of Chemistry for Materials, Faculty of Engineering, Mie University, Tsu Mie 514-8507, Japan}
\date{\today}
\maketitle
\begin{abstract}
Conformations of a single semiflexible polymer chain dissolved in a low molecular weight liquid crystalline solvents (nematogens) are examined by using a mean field theory. We takes into account a stiffness and partial orientational ordering of the polymer. As a result of an anisotropic coupling between the polymer and nematogen, we predict a discontinuous (or continuous) phase transition from a condensed-rodlike conformation to a swollen-one of the polymer chain, depending on the stiffness of the polymer. We also discuss the effects of the nematic interaction between polymer segments.
\end{abstract}

\narrowtext

Mesomorphic mixtures consisting of polymers and low molecular weight liquid crystals (nematogens) are of current interest for fundamental scientific reasons and for many technological applications in electro-optical devices and high modulus fibers. The performances of these systems are related to a conformation of a polymer in a liquid crystal phase. One of the fundamental problem is how the polymer in a nematic phase interacts with a nematic field surrounding the polymer. Polymer chains have the variety of their stiffness and so when the polymer is mixed with the nematogens we can expect various conformations.

Mixtures of a flexible polymer and a nematogen show a macroscopic phase separation between an isotropic and a nematic phase below the nematic-isotopic transition (NIT) temperature of the pure nematogen\cite{kronberg78,dubaut80,brochard84,shen95,loudet00}. Flexible polymers present a weak anisotropy in a nematic phase\cite{dubault85}. In contrast, liquid crystalline polymers, or stiffer polymers, have good miscibility with nematogens\cite{shibaev93} due to the strong anisotropic coupling between the polymer and the nematogen. Anisotropy of the conformation for liquid crystalline polymers has been experimentally\cite{shibaev93,dallest88,chiang97} and theoretically\cite{ten83,warner85,flory78,odijk96,carri98,peter94,matsuyama99} studied in melt and in dilute nematic solutions. It is now important to consider the conformation of a polymer chain with various degrees of stiffness dissolved in nematogens. Recently we presented a mean field theory to describe partial orientational ordering (induced rigidity) of semiflexible polymers dissolved in nematogens and showed various phase diagrams for the mixtures\cite{matsuyama99}. 

In this Rapid Communication we theoretically study the conformation of a semiflexible polymer dissolved in nematic solvents by combining the previous model\cite{matsuyama99} with an elastic free energy of the chain\cite{matsuyama01}. We show a discontinuous (or continuous) conformational transition between two different nematic states, depending on the stiffness of the polymer.

Consider a single linear polymer chain dissolved in nematogens. In order to take into account the stiffness of the polymer, we here assume that two neighboring bonds on the polymer chain have either bent (gauche state) or straightened (trans state) conformations\cite{flory56a,matsuyama97}. Hereafter we refer the segments in straightened bonds  as "rigid" segments. 

Let $V=R^3$ be the volume of the region occupied by a polymer, $n$ be the number of segments on the polymer chain and $n_l$ be the axial ratio of the nematogen. The volume fraction of the polymer in the volume $V$ is given by 
$
\phi=a^3n/V,
$
where $a^3$ is the volume of an unit segment. To derive an equilibrium conformation of the polymer, we consider thermodynamics of our systems. The free energy density of our system can be given by
\begin{equation}
f=f_{el}+f_{bent}+f_{mix}+f_{nem}.
\label{eq:201}
\end{equation}
The first term shows the elastic free energy due to the deformation of the polymer chain. Let $R_z$ be the length of the polymer along the direction z of the nematic director and $R_x$ be the length along the perpendicular direction ($R^3=R_x^2R_z$). Combining the classical elastic free energy obtained by Flory\cite{flory53} with the freely jointed rod model\cite{matsuyama01}, the elastic free energy is given as a function of $\phi$ and an orientational order parameter $S_r$ of the rigid segments\cite{matsuyama01}:
\begin{equation}
\beta f_{el}=\frac{3}{2n}{\Big[}{\Big(}\frac{\phi}{nA} {\Big)}^{1/3}+\frac{\phi}{3}\ln A
-\phi(1-\frac{2}{3}\ln \sqrt{n}\phi){\Big ]},
\label{eq:202}
\end{equation}
where $A\equiv(1+2S_r)(1-S_r)^2$ and $\beta\equiv k_BT$, $T$ is the absolute temperature, $k_B$ is the Boltzmann constant.

The second term shows the free energy change needed to straighten bent bonds on the polymer and is given by\cite{flory56a,matsuyama97}
\begin{equation}
\beta f_{bent}=n_r(\beta f_0)-S_{comb}/k_B,
\label{eq:203}
\end{equation}
where $f_0$ is the local free energy difference between the bent and straightened conformations, and $n_r$ shows the number of the rigid segments on the polymer. The second term in Eq. (\ref{eq:203}) is the combinatorial entropy related to the number of ways to select $n_r$ rigid segments out of the $n$ segments on the polymer and is given $S_{comb}/k_B=\ln(n!/[n_r!(n-n_r)!])=-n[x\ln x+(1-x)\ln (1-x)]$, where we used the Stirling's approximation and $x(\equiv n_r/n)$ shows the fraction of the rigid segments on the polymer. The volume fraction $\phi_r$ of the rigid segments on the polymer chain is given by $\phi_r=x\phi$. The third term in Eq. (\ref{eq:201}) is the free energy of the isotropic mixing for the polymer and nematogens. According to the Flory theory\cite{flory53}, the free energy is given by
\begin{equation}
\beta f_{mix}=\frac{1-\phi}{n_{l}}\ln (1-\phi)+\chi\phi(1-\phi),
\label{eq:204}
\end{equation}
where $\chi(\equiv U_0/k_BT)$ is the Flory-Huggins interaction parameter related to the isotropic dispersion interactions between unlike molecular species.

The last term in Eq. (\ref{eq:201}) shows the free energy for the nematic ordering. On the basis of the Maier-Saupe model\cite{maier59,degennes93} for orientational dependent-attractive interactions, the free energy of the nematic ordering is given by
\begin{eqnarray}
\beta f_{nem}&=&\frac{1-\phi}{n_{l}}\int f_{l}(\theta)\ln 4\pi f_{l}(\theta) d\Omega \nonumber \\
&&
+\frac{\phi_r}{n_r}\int f_r(\theta)\ln 4\pi f_r(\theta) d\Omega
-\frac{1}{2}\nu_{ll}S_{l}^2(1-\phi)^2  \nonumber \\
&&
-\nu_{lr}S_{l}S_r(1-\phi)\phi_r-\frac{1}{2}\nu_{rr}S_{r}^2\phi_r^2
\label{eq:205}
\end{eqnarray}
where $d\Omega\equiv 2\pi \sin\theta d\theta$, $\theta$ is the angle between the nematogen and the director of the orienting field. The $\nu_{ll}$ shows the orientational dependent (Maier-Saupe) interactions between the nematogens, $\nu_{lr}$ is that between the nematogen and the rigid segment on the polymer, and $\nu_{rr}$ is that between rigid segments. The $f_{l}(\theta)$ and $f_r(\theta)$ show the orientational distribution functions of the nematogens and that of the rigid segments on the polymer, respectively. The orientational order parameter $S_{l}$ of the nematogens and that $S_r$ of the rigid segments is given by
\begin{equation}
S_{i}=\int P_2(\cos\theta) f_{i}(\theta)d\Omega,
\label{eq:206}
\end{equation}
$i=l,r$, where $P_2(\cos\theta)\equiv 3(\cos^2\theta-1/3)/2$. 

The orientational distribution functions $f_{l}(\theta)$  and $f_r(\theta)$ are determined by the free energy (\ref{eq:201}) with respect to these functions: 
$(\partial F_{nem}/\partial f_{i}(\theta))_{x,f_j}=0,$
under the normalization conditions
$\int f_i(\theta)d\Omega=1.$ We then obtain the distribution function:
\begin{equation}
f_{i}(\theta)=\frac{1}{Z_{i}}\exp[\eta_{i}P_2(\cos\theta)],
\label{eq:207}
\end{equation}
\begin{equation}
\eta_{l}\equiv n_{l}{\Big[}\nu_{ll}S_{l}(1-\phi)+\nu_{lr}S_rx\phi{\Big]},
\label{eq:208}
\end{equation}
\begin{equation}
\eta_r\equiv n {\Big[}x\nu_{lr}S_{l}(1-\phi)+x^2\nu_{rr}S_r\phi-E(\phi,S_r){\Big]},
 \label{eq:209}
\end{equation}
where 
\begin{equation}
E\equiv \frac{3S_r}{n\phi(1+2S_r)(1-S_r)}{\Big[}{\Big(}\frac{\phi}{nA} {\Big)}^{1/3}-\phi{\Big]}.
\label{eq:210}
\end{equation}
The constants $Z_i$ ($i={l},r$) are determined by the normalization condition as
$Z_{i}=2\pi I_0[\eta_{i}],$
where the function $I_0[\eta_i]$ is defined as
\begin{equation}
I_q[\eta_i]\equiv\int_0^1{\Big[}P_2(\cos\theta){\Big]}^q\exp{\Big[}\eta_iP_2(\cos\theta){\Big]}d(\cos\theta),
\label{eq:211}
\end{equation}
$q=0,1,2,\cdots.$ Substituting Eq. (\ref{eq:207}) into (\ref{eq:206}), we obtain two self-consistency equations for the two order parameters $S_{l}$ and $S_r$:
\begin{equation}
S_{i}=I_1[\eta_{i}]/I_0[\eta_{i}].
\label{eq:212}
\end{equation}
The orientational order parameter of the polymer chain is given by $S_p=xS_r$.

The fraction $x$ of the rigid segments on the polymer is determined by minimizing the free energy (\ref{eq:201}) with respect to $x$: $(\partial F/\partial x)_{S_{l},S_r}=0$. This yields
\begin{equation}
\nu_{rr}\phi S_r^2x+\nu_{lr}S_rS_l(1-\phi)-\ln\frac{x}{(1-x)K}=0
\label{eq:213}
\end{equation}
where $K\equiv\exp(-\beta f_0)$.
By solving the coupled equations (\ref{eq:212}) and (\ref{eq:213}), we can obtain the values of the two order parameters $S_{l}$, $S_r$, and the fraction $x$ of the rigid segments as a function of temperature and concentration $\phi$. 

The chemical potential $\mu_l(\phi)$ of the nematogen inside the volume $V$ occupied by the polymer can be calculated by 
$\mu_l=f-\phi(df/f\phi)$:
\begin{eqnarray}
\beta \mu_l(\phi)&=&\frac{1}{n}{\Big[}\Big(\frac{\phi}{nA}\Big)^{1/3}-\phi{\Big]}
+\frac{1}{n_l}\ln(1-\phi)+\frac{\phi}{n}+\chi\phi^2
\nonumber \\
&&
+\frac{1}{2}\nu_{ll}S_{l}^2(1-\phi)^2
+ \nu_{lr}S_{l}S_rx\phi(1-\phi)
\nonumber \\
&&
+\frac{1}{2}\nu_{rr}S_{r}^2x^2\phi^2
-\frac{1}{n_l}\ln I_0[\eta_l],
\label{eq:214}
\end{eqnarray}
and the chemical potential $\mu_l^{\circ}(S_b)$ of the nematogen  outside the polymer is given by substituting $\phi=0$ into $\mu_l(\phi=0)$. The orientational order parameter $S_b$ of the bulk nematogens outside the polymer is determined by the self-consistency equation: $S_b=I_1[\eta_b]/I_0[\eta_b],$ where $\eta_b\equiv n_l\nu_{ll}S_b$. The equilibrium concentration $\phi$ of the polymer can be determined from the balance among the nematogens existing outside and inside the polymer:
$
\mu_l(\phi)=\mu_l^{\circ}(S_b)
$

In our numerical calculations, we further split the local free energy difference $f_0$ in Eq. (\ref{eq:203}) into two parts:
$
f_0=\epsilon_0-Ts_0,
$
where $s_0(=k_B\ln\omega)$ is the local entropy loss and $\epsilon_0(<0)$ is the energy change needed to straighten a bent bond. We then obtain $K=\omega\exp(-\beta\epsilon_0)$. The anisotropic interaction parameter $\nu_{ll}$ is given to be inversely proportional to temperature:\cite{degennes93}
$
\nu_{ll}=U_a/k_BT.
$
We define the dimensionless nematic interaction parameter
$
\alpha\equiv \nu_{ll}/\chi
$ 
and the stiffness parameter $\epsilon\equiv -\beta\epsilon_0/\nu_{ll}$ of a polymer. The larger values of $\epsilon$ correspond to the stiffer chains. The most flexible polymer chain is realized when $\epsilon=0$.  We also put $b=\nu_{lr}/\nu_{ll}$ and $c=\nu_{rr}/\nu_{ll}$, where $b$ and $c$ are constants. In the following calculations we use $\alpha =5$, $n_l=2$, $n=100$, $\omega=0.025$ , and $b=1$ for a typical example. When $\nu/\chi=5$, or for the larger values of $\nu_{ll}/\chi$, the nematogen behaves as a good solvent for the polymer because the value of the interaction parameter ($\chi$) between the polymer and the solvent molecule is small\cite{matsuyama99}. However, the anisotropic interaction between the polymer and the nematogen exists.

\begin{figure}
\includegraphics*[height=7cm]{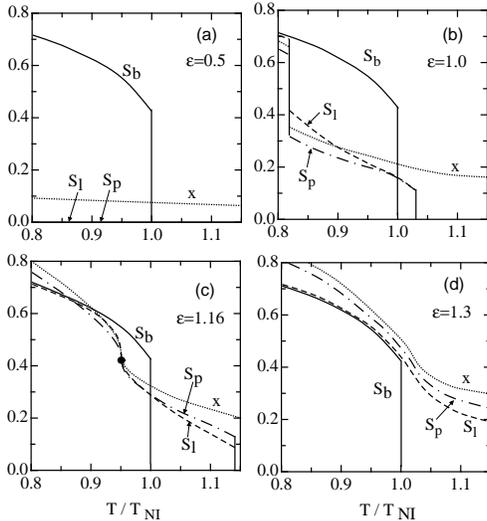}
\caption{Orientational order parameters and the fraction $x$ of the rigid segments on the polymer chain plotted against the reduced temperature $T/T_{NI}$, where $T_{NI}$ is the NIT temperature of the pure nematogen outside the polymer. The value of stiffness parameter $\epsilon$ of the polymer is changed from (a) to (d).}
\label{fig1}
\end{figure}

\begin{figure}
\includegraphics*[height=6cm]{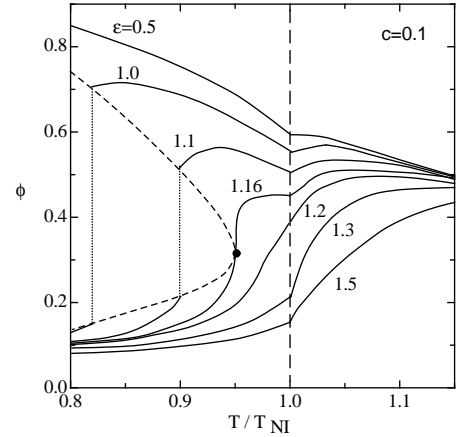}
\caption{Equilibrium volume fraction $\phi$ (swelling curve) of the polymer plotted against the reduced temperature for $c=0.1$. The stiffness ($\epsilon$) of the polymer is changed.}
\label{fig2}
\end{figure}

\begin{figure}
\includegraphics*[height=6cm]{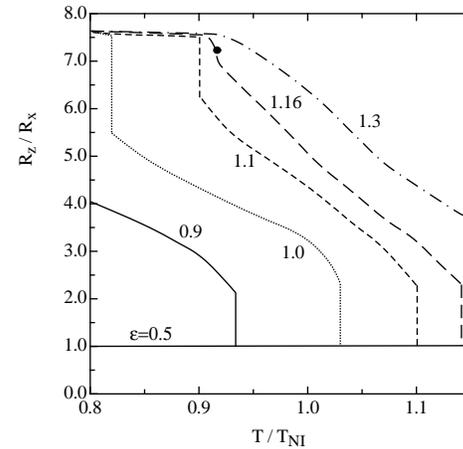}
\caption{Anisotropy $R_z/R_x$ plotted against the temperature, where $R_z$ ($R_x$) is the projection of the radius of gyration parallel (perpendicular) to the nematic director.}
\label{fig3}
\end{figure}

Figures \ref{fig1}-\ref{fig3} show the results calculated for $c=0.1$ with a weak nematic interaction $\nu_{rr}$. Figure \ref{fig1} shows the orientational order parameters and the fraction $x$ of the rigid segments on the polymer plotted against the reduced temperature $T/T_{NI}$, where $T_{NI}$ is the NIT temperature of the pure nematogen outside the polymer. The value of stiffness parameter $\epsilon$ of the polymer is changed from (a) to (d). The solid curve refers to the order parameter $S_{b}$ of the nematogen outside the polymer and the dash-dotted line shows the order parameter $S_p$ of the polymer. The short-dashed line shows the order parameter $S_l$ of the nematogen inside the polymer and the dotted line corresponds to the fraction $x$ of the rigid segments. For $\epsilon=0.5$, or a flexible polymer, the polymer is in an isotropic state for all temperatures and there is no anisotropic coupling between the polymer and nematogen. When $\epsilon=1.0$, we find two phase transition temperatures: one is the temperature $T_{NI}^p$ at high temperatures where the NIT of the polymer takes place and the other is the $T_{NN}$ at low temperatures where the first-order phase transition between two different nematic states takes place. At the nematic state of the high temperature side ($T_{NN}<T<T_{NI}^p$), the fraction $x$ of rigid segments is small. The polymer and the nematogen inside the polymer are slightly ordered (weakly ordered nematic phase). In another nematic phase at $T<T_{NN}$, the value of the fraction $x$ is large and the polymer and the nematogen are strongly ordered (strongly ordered nematic phase). As increasing the stiffness,  $T_{NI}^p$ and $T_{NN}$ move to higher temperatures and a critical point (closed circle) for $T_{NN}$ appears at $\epsilon=1.16$. 

Figure \ref{fig2} shows the equilibrium volume fraction $\phi$ (swelling curve) of the polymer plotted against the reduced temperature. The stiffness ($\epsilon$) of the polymer is changed.  When $\epsilon=0.5$, as decreasing temperature, the polymer is continuously condensed (or the volume fraction of the polymer is increased). The swelling curve has as kink at $T_{NI}$. For stiffer polymers, we find the first-order phase transition between a condensed conformation and a swollen one at $T_{NN}$. As shown in Fig \ref{fig3}, the polymer is elongated along the nematic field and has a rodlike conformation. We then find two different types in the rodlike conformation of the polymer chain: one is the swollen-rodlike conformation at $T<T_{NN}$ and the other is the condensed-rodlike conformation at $T_{NN}<T<T_{NI}^p$. Above the critical stiffness, the polymer is continuously swollen with decreasing temperature. At the lower temperatures, the polymer is swollen because the anisotropic interaction between the polymer and nematogen prevails.

Figure \ref{fig3} shows the anisotropy $R_z/R_x$ of the polymer chain plotted against the reduced temperature for various values of $\epsilon$. When $\epsilon=0.5$, we have $R_z=R_x$ and the polymer has a spherical (coil) conformation. For $\epsilon=1.0$, as decreasing temperature, the polymer is changed from a coil to a rodlike conformation at $T_{NI}^p$. The concentration $\phi$ of the polymer segments is continuously changed at $T_{NI}^p$ (see Fig. \ref{fig2}). This spontaneous change in the polymer conformation at the transition $T_{NI}^p$ has been reported\cite{dallest88,ten83,warner85}. Further decreasing temperatures, we find the phase transition between the two different rodlike conformations at $T_{NN}$. Above the critical stiffness, the polymer continuously changed from the condensed-rodlike conformation to the swollen-one with decreasing temperature.

\begin{figure}
\includegraphics*[height=6cm]{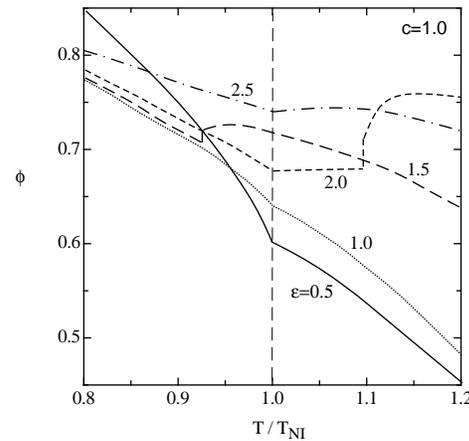}
\caption{Swelling curve of the polymer for $c=1.0$. The stiffness ($\epsilon$) of the polymer is changed.}
\label{fig4}
\end{figure}

When the anisotropic (attractive) interaction $\nu_{rr}$ between the polymer segments is strong, the swollen-rodlike conformation at lower temperatures disappears. Figure \ref{fig4} shows the swelling curve of a polymer with $c=1.0$. The stiffness ($\epsilon$) of the polymer is changed. As decreasing temperature, the polymer is condensed with a rodlike conformation because the anisotropic interaction between the polymer segments prevails at lower temperatures. When $\epsilon=1.5$ and $2.0$, we have a small jump in the swelling curve at $T_{NN}$\cite{matsuyama}. The anisotropic  interaction becomes an important factor on the conformation of a semiflexible polymer. 

In conclusion, we have predicted two different rodlike conformations of a polymer chain in a nematic solvent. The phase transition from a condensed-rodlike conformation to a swollen-one is strongly affected by the stiffness of a polymer and the anisotropic interaction between polymer segments. The concept of the two different rodlike conformations is important to the modification of the mechanical and viscoelastic properties of liquid crystals by the dissolved polymers. The results will also be important to the conformation of DNA chains\cite{odijk96} and the volume phase transition of nematic gels\cite{urayama02} in nematic solvents.

This work was supported by a Grant-in-Aid from the Ministry of
Education, Culture, Sports, and Science and Technology, Japan.

\end{document}